\documentclass[conference]{IEEEtran}
\IEEEoverridecommandlockouts
\usepackage{cite}
\usepackage{amsmath,amssymb,amsfonts}
\usepackage{algorithmic}
\usepackage{graphicx}
\usepackage{float}
\usepackage{textcomp}
\usepackage{xcolor}
\def\BibTeX{{\rm B\kern-.05em{\sc i\kern-.005em b}\kern-.08em
    T\kern-.1667em\lower.7ex\hbox{E}\kern-.125emX}}
\begin{document}

\title{Dynamic Analysis of electrostatically actuated fixed-fixed microbeams with midplane-stretching effects
\\
}

\author{\IEEEauthorblockN{1\textsuperscript{st} Ganesh Iyer}
\IEEEauthorblockA{\textit{Department of Mechanical Engineering} \\
\textit{Indian Institute of Technology, Bombay}\\
% Mumbai, India\\
% 210100059@iitb.ac.in
}
\and
\IEEEauthorblockN{2\textsuperscript{nd} MM Joglekar}
\IEEEauthorblockA{\textit{Department of Mechanical Engineering } \\
\textit{Indian Institute of Technology, Roorkee}\\
% Mumbai, India \\
% pawaskar@iitb.ac.in
}
\and
\IEEEauthorblockN{3\textsuperscript{nd} D.N. Pawaskar}
\IEEEauthorblockA{\textit{Department of Mechanical Engineering } \\
\textit{Indian Institute of Technology, Bombay}\\
% Mumbai, India \\
% pawaskar@iitb.ac.in
}
}

\maketitle

\begin{abstract}
We explore the undamped response of step-voltage driven parallel plates and fixed-fixed microbeams. We consider a third order correction in the parallel plate system as a reduced order system to model mid-plane stretching in fixed-fixed beams. In both cases, we employ energy methods to determine the threshold values displacement and voltage. This is the point beyond which there is no oscillatory motion. The dynamic pull-in parameters of the corresponding microactuator model are identified as these critical values.  We thoroughly analyze the effects of mid-plane stretching in fixed-fixed microbeams. Additionally, the phase portraits for these actuators are also thoroughly examined. Empirical relations for the oscillation period/switching time for these microactutators are also developed. These results correlate to a high accuracy to previously published results. 
\vspace{2mm}
\end{abstract}


%\section{Introduction}

\section{Dynamic Analysis of Non-linear Parallel Plates Model}
In this section, we conduct dynamic analysis for a standard reduced order parallel plates model. However, we include a cubic order correction in the spring force to model mid-plane stretching, which is prominent in fixed-fixed beams. Many researchers have analyzed the linear suspension parallel plates model, using techniques such as the energy technique, hamiltonian approach, and numerical integration of the momentum equation. We use the energy technique for the subsequent analysis.

\begin{figure}[htbp]
\centerline{\includegraphics[width=0.5\textwidth]{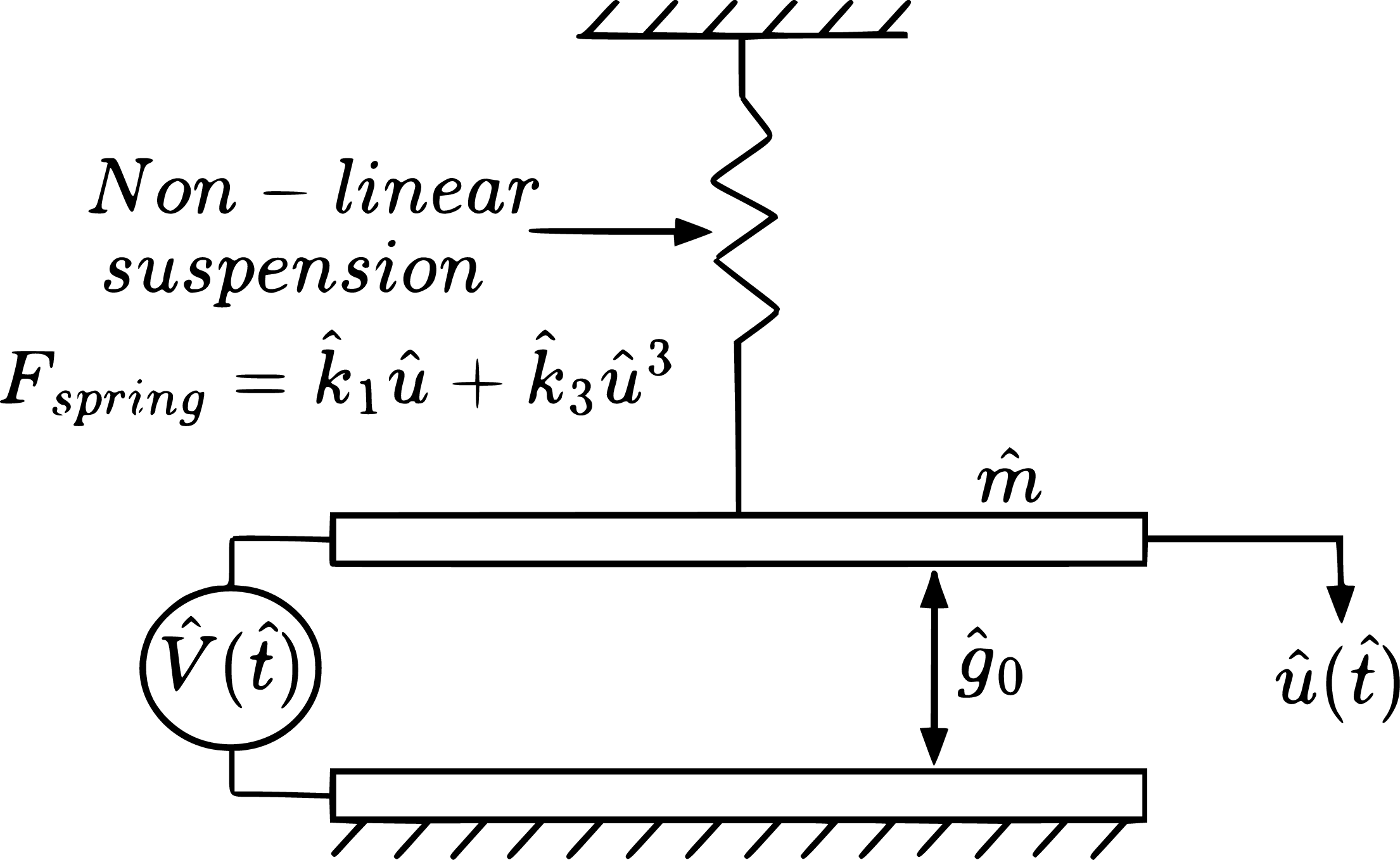}}
\caption{Parallel plates model with non-linear suspension}
\label{fig:Parallel_plate}
\end{figure}
The schematic for our parallel plate system is shown in 1. Throughout our discussion, $(\hat{.})$ would refer to physical dimensional quantities. The top movable electrode has a mass $\hat{m}$, and it is attached to a non-linear spring with constants $\hat{k}_1$ and $\hat{k}_3$. The initial separation between the electrodes is $\hat{g}_0$. Permittivity of vacuum is $\hat{\epsilon}_0$ and $\hat{A}$ is the area of overlap between the two electrodes. $\hat{V}(\hat{t}) = \hat{V}H(\hat{t}) $ is the applied voltage step input, where $H(\hat{t})$ is the heaviside function. $\hat{u}(\hat{t})$ is the time-dependent displacement of the movable electrode.
\par
Neglecting energy dissipations, the dynamical equation of the system is given as
\begin{equation}
     \hat{m}\frac{d^2\hat{u}}{d\hat{t}^2} + \hat{k_1}\hat{u} + \hat{k_3}\hat{u}^3 = \frac{\hat{\epsilon_o}\hat{A}\hat{V}^2}{2(\hat{g_o}-\hat{u})^2}
\end{equation}
We use the following dimensionless quantities to generalize our subsequent discussion :
\begin{equation*}
    u(t) = \frac{\hat{u}(\hat{t})}{\hat{g_0}}~, ~    V = \sqrt{\frac{\hat{\epsilon_o}\hat{A}\hat{V}^2}{\hat{k_1}\hat{g_0}^3}}~, ~
    t = \left(\sqrt{\frac{\hat{k_1}}{\hat{m}}}\right)\hat{t}~, ~
    \alpha = \frac{\hat{k_3}\hat{g_o}^2}{\hat{k_1}}
\end{equation*}
Here u, V, and t represent dimensonless displacement, voltage and time resepctively, and $\alpha$ will be called the stiffening parameter. This gives us the following non-dimensional form
\begin{equation}
    \frac{d^2u}{dt^2} + u + \alpha u^3 = \frac{V^2}{2(1-u)^2}
\end{equation}
We use the following two initial conditions -
\begin{equation}
    u|_{t=0} = 0~,~~~~ \dot{u}|_{t=0} = 0
\end{equation}

\subsection{Determination of Dynamic pull-in parameters}
Several past investigations reveal that at the critical pull-in point, velocity of the moving electrode is 0. We use this fact in our energy method, writing a balance between the stored potential energy of the spring and the supplied electrostatic energy, as shown below -
\begin{equation}
    \int_{0}^{\tilde u} u + \alpha u^3 du = \int_{0}^{\tilde u} \frac{\tilde V ^2 du}{2(1-u)^2}
\end{equation}
$\tilde u$ represents the amplitude for a given applied voltage $\tilde V$. Simplifying the above equation, we obtain 
\begin{equation}
    \tilde V = \sqrt{\left(\tilde u + \frac{\alpha \tilde u ^3}{2}\right)(1 - \tilde u)}
\end{equation}
We now use the condition of instability, which is 
\begin{equation}
    \frac{d\tilde V}{d\tilde u} = 0
\end{equation}
Using numerical root-finding methods, we find the pull-in displacement for a given alpha, and thus the pull-in voltage. We thus get the dynamic pull-in parameters as functions of the stiffening parameter $\alpha$. The variation of the pull-in displacements, along with their polynomial fits is shown below. 
\begin{center}
   \includegraphics[scale=0.6]{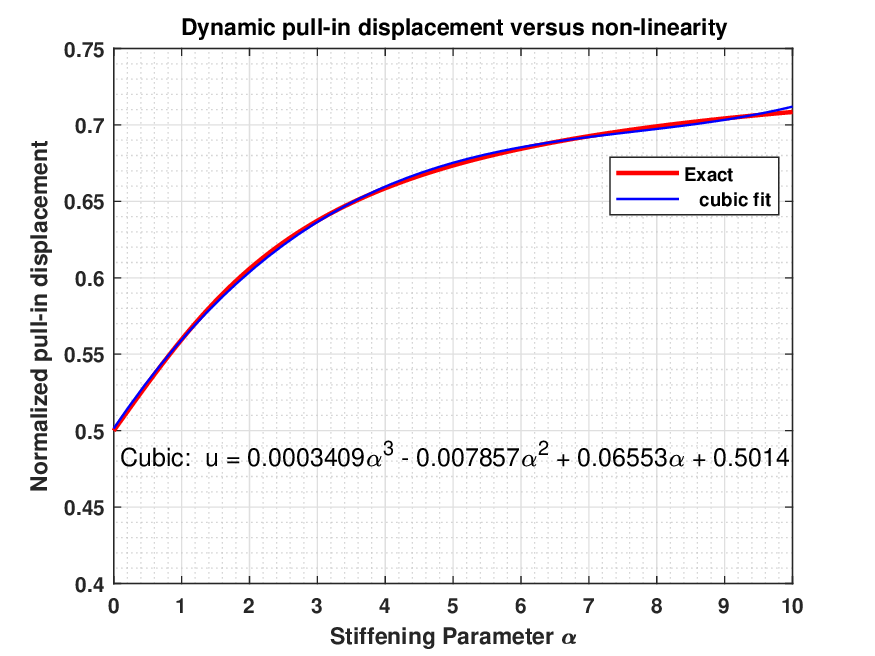}
\end{center}
\begin{center}
   \includegraphics[scale=0.6]{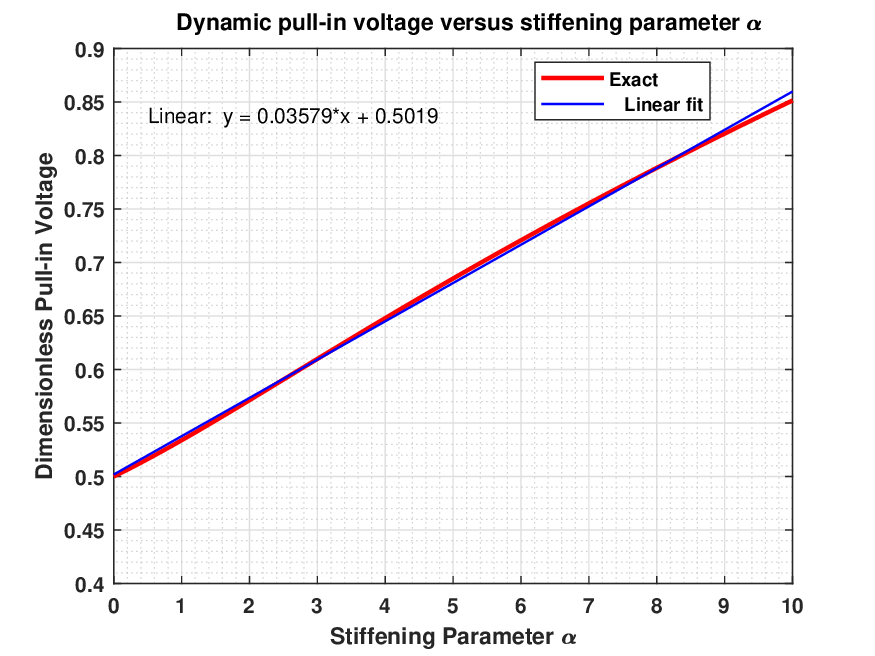}
\end{center}
We observe that both pull-in displacement as well as pull-in voltage increases monotonically with the stiffening parameter $\alpha$. Another interesting observation to note is that the pull-in voltage is almost linear in vriation, for the given $\alpha$ range.
\subsection{Phase Plane Analysis}
We find the numerical value of velocity at any instant in time using energy conservation, i.e sum of the kinetic energy of the electrode and elastic energy in the spring is the supplied electrostatic energy. We can write it mathematically as shown below,
\begin{equation}
    \frac{\dot u^2}{2} + \int_{0}^{u} x + \alpha x^3 dx =  \frac{V^2 u}{2(1-u)}
\end{equation}
Using this energy conservation, we get dimensionless velocity as a function of dimensionless displacement, V and $\alpha$.
\begin{equation}
    \dot u = \sqrt{\frac{V^2u}{(1-u)}-u^2 - \alpha \frac{u^4}{2}}
\end{equation}
We make the phase plot for a given dimensionless applied voltage of 0.7, to analyze the effect of the stretching parameter $\alpha$ on the phase plot. We find that for this applied voltage to be pull-in, the value of $\alpha$ is 5.41 from the pull-in parameter calculation.
\begin{center}
   \includegraphics[scale=0.7]{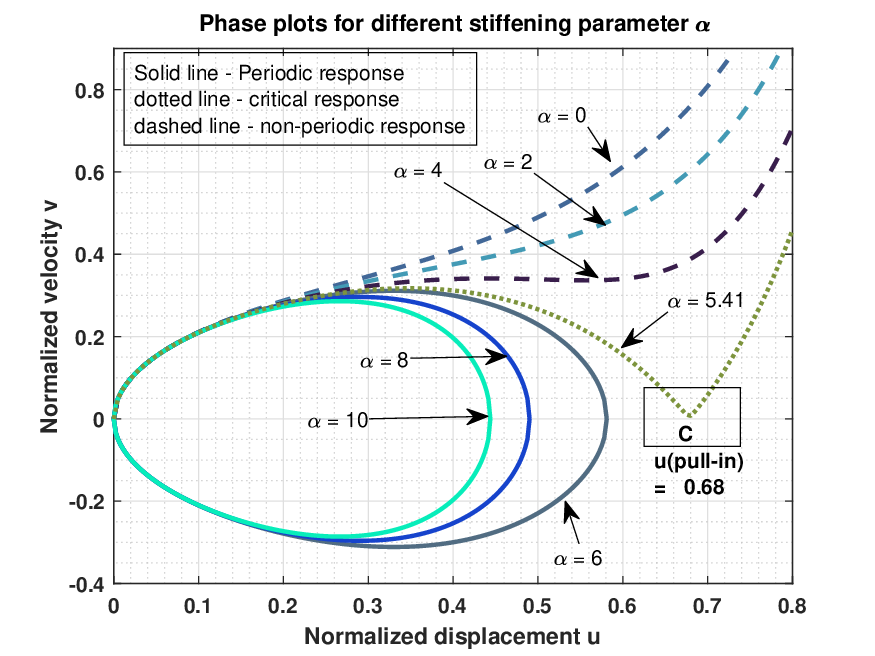}
\end{center}

As the pull-in voltage consistently rises with $\alpha$, it is evident that for values of $\alpha$ below 5.41, the condition V=0.7 surpasses the threshold of 0.7, resulting in a non-periodic response. Conversely, for higher $\alpha$ values, a periodic response ensues. Therefore, the progressive increase in $\alpha$ distinctly transforms the response nature from non-periodic to periodic.

A notable aspect worth highlighting is the compression or flattening of the ellipse-like periodic orbits at the extreme displacement end. This phenomenon is a direct consequence of the reduction in the plate's maximum amplitude.

In the subsequent section, our analysis will focus on fixed-fixed beams. Within this context, we aim to provide a theoretical explication of these observations, specifically addressing the impact of the stretching effect. This effect has been mathematically incorporated through the stiffening parameter $\alpha$ within the framework of this non-linear spring and plate model.
\subsection{Time Measures for the non-linear model}
We can use the velocity expression obtained to form the phase plots, and using it, find the oscillation period when the response is periodic, and the switching time if the response is non-periodic. If the maximum amplitude of normalized displacment is $\tilde u$, the oscillation period is written as 
\begin{equation}
     T_p = 2\int_0^{\tilde u} \frac{du}{(du/dt)}  = 2\int_0^{\tilde u} \frac{du}{\sqrt{\frac{V^2u}{(1-u)}-u^2 - \alpha \frac{u^4}{2}}}
\end{equation}
For non-periodic response, the maximum amplitude is 1(till plate hits the other base electrode). So, switching time can be written as 
\begin{equation}
    T_{np} = \int_0^{1} \frac{du}{(du/dt)} \\
    = \int_0^{1} \frac{du}{\sqrt{\frac{V^2u}{(1-u)}-u^2 - \alpha \frac{u^4}{2}}}
\end{equation}
We plot the oscillation frequencies for the periodic response, namely $2\pi/T_p$, and compare it with analogous graph for fixed fixed beams, to analyze the effect of stretching modelled as stiffening parameter $\alpha$.
\begin{center}
   \includegraphics[scale=0.6]{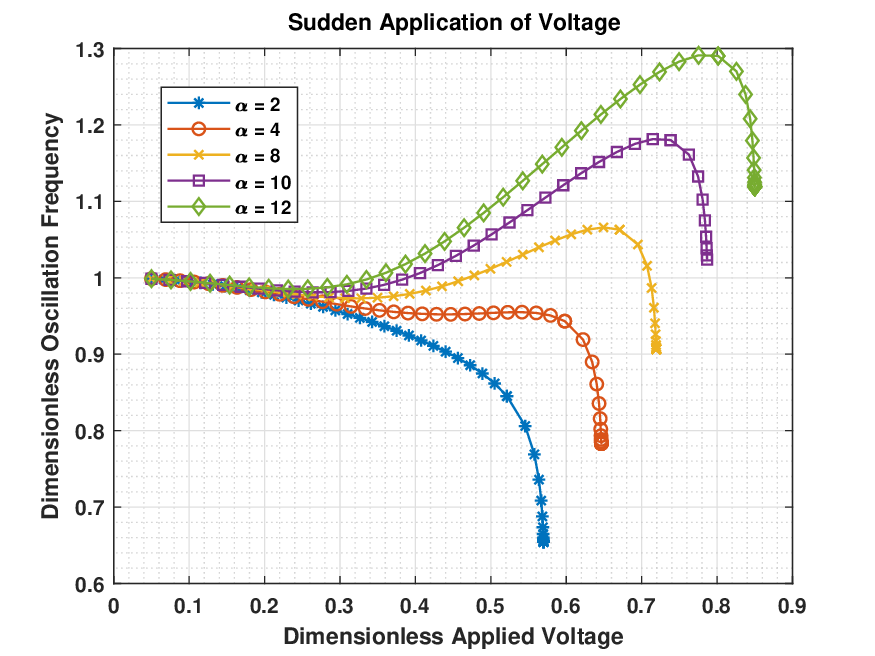}
\end{center}
As the applied voltage tends to the value of the pull-in voltage for the respective $\alpha$, the frequency drops to 0 with an extremely high slope, because it is the transition between periodic to non-periodic response. To capture the high slope, we have tried to include many data points close to the pull-in voltage. \\
Some interesting observations that can be drawn from here are that for smaller $\alpha$, there is a monotonic decrease in oscillation frequency, while for higher zeta, it rises to a maximum value and then it starts to drop. We will come back to explain this effect in the analysis of fixed-fixed beams.

\section{Dynamic Analysis of fixed fixed beams}
Fig 2 shows a schematic of an electrostatically actuated slender, prismatic fixed-fixed microbeam made up of linear elastic material having Young’s modulus $\hat{E}$, Poisson’s ratio $ \nu$ and density $\hat{\rho}$. $\hat{L}$, $\hat{b}$ and $\hat{}$h denote length, breadth and
thickness of the microbeam respectively. $\hat{I}$
is the moment of inertia of the cross section. Initially the two electrodes are separated by a
distance equal to $\hat{g}_0$ and the permittivity of the free space is $\hat{\epsilon}_0$. The effective Young’s modulus of the microbeam material is denoted by $\hat{E^*}$ . For narrow microbeams $(\hat{b}<5\hat{h})$, $\hat{E^*} =
\hat{E}$, while for
wide microbeams $(\hat{b}
\geq 5\hat{h})$, $\hat{E^*} = \hat{E}/(1-\nu^2)$. We are neglecting fringing effects of the elctrostatic field since it is a second order effect, and comes into role only for extremely thin wire-like beams.
\begin{figure}[htbp]
\centerline{\includegraphics[width=0.4\textwidth]{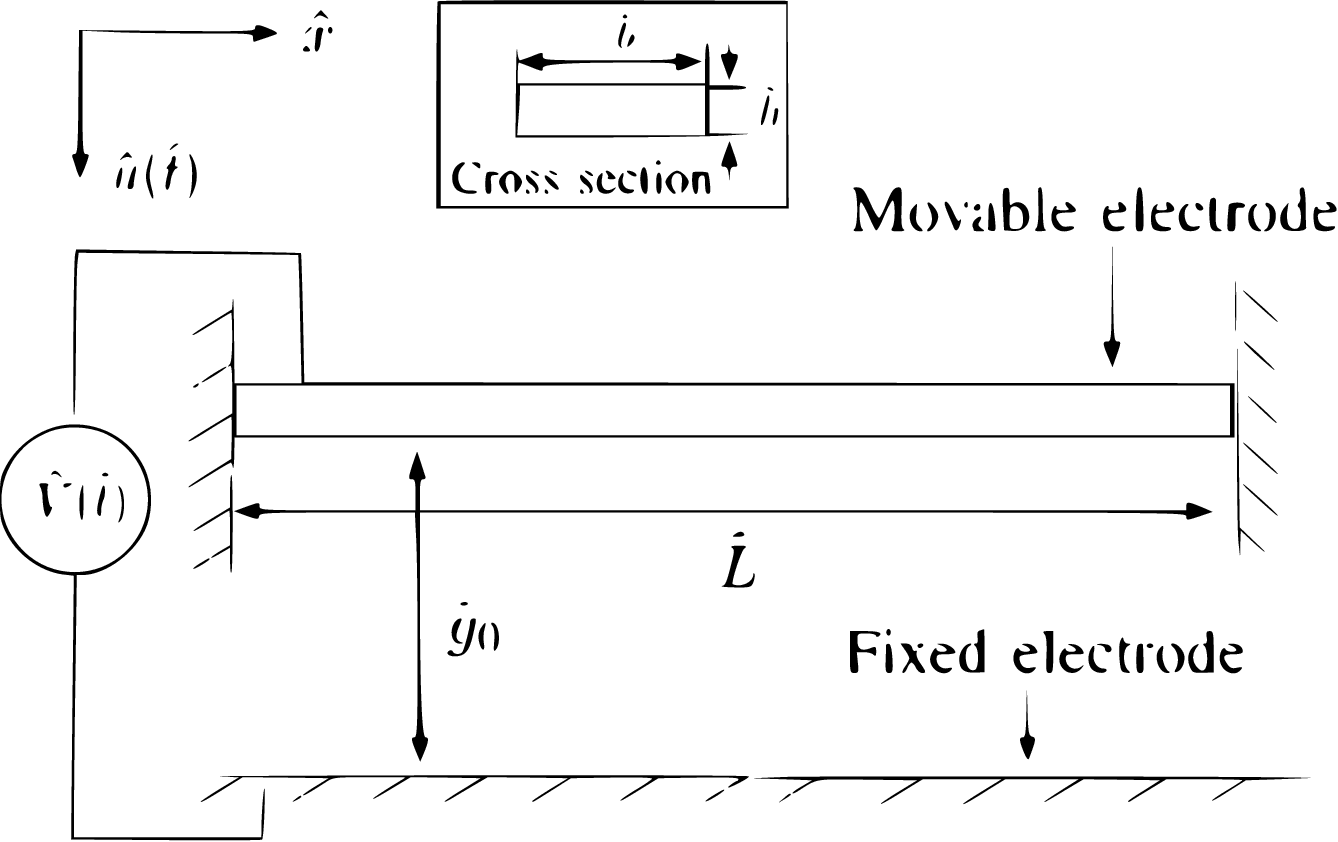}}
\caption{Schematic of an electrostatically actuated fixed–fixed beam}
\end{figure}
We now analyze the response of the microbeam for sudden application of voltage, i.e $\hat{V}(\hat{t}) = \hat{V} \hat{H}(\hat{t})$, where $\hat{V}$ is the magnitude of the
voltage signal and $\hat{H}(\hat{t})$ is the Heaviside step function. The governing differential equation of motion of the microbeam under the
application of the step-voltage is written as
\begin{equation} 
\begin{split}
& \hat{\rho}\hat{b}\hat{h}\frac{\partial ^2 \hat{u}(\hat{x},\hat{t})}{\partial \hat{t}^2} + \hat{E}^*\hat{I}\frac{\partial ^4 \hat{u}(\hat{x},\hat{t})}{\partial \hat{x}^4} - 
    \left( \frac{\hat{b}\hat{h}\hat{E}}{2\hat{L}}\int^{\hat{L}}_0\left(\frac{\partial \hat{u}(\hat{x},\hat{t})}{\partial \hat{x}}\right)^2 d\hat{x}\right)\frac{\partial ^2 \hat{u}}{\partial \hat{x}^2} \\
& = \frac{\hat{\epsilon_o}\hat{b}(\hat{V}(\hat{t}))^2}{2(\hat{g_o} - \hat{u}(\hat{x},\hat{t}))^2}
\end{split}
\end{equation}
We define the following dimensional quantities to non-dimensionalize our equations for subsequent analysis:
\begin{equation}
    \begin{split}
        & u(x,t) = \frac{\hat{u}(\hat{x},\hat{t})}{\hat{g}_0},~~ x = \frac{\hat{x}}{\hat{L}},~~  t = \frac{\hat{t}\sqrt{\hat{E^*}\hat{I}}}{\sqrt{\hat{\rho}\hat{b}\hat{h}\hat{L^4}}}, ~~
        V =  \sqrt{\frac{\hat{\epsilon}_0\hat{b}\hat{L^4}\hat{V}^2}{2\hat{E^*}\hat{I}\hat{g}^3_0}} \\
        & M = \frac{\hat{E}}{\hat{E}^*} ,~~ \xi = \left(\frac{\hat{g_o}}{\hat{h}}\right)^2
    \end{split}
\end{equation}
Here, we call the parameter $\xi$ as the stretching ratio, and M as the modulus ratio.
Thus, Eq.(11) is reduced to the following non-dimensional form:
\begin{equation}
\begin{split}
    & \frac{\partial ^2 u(x,t)}{\partial t^2} + \frac{\partial ^4 u(x,t)}{\partial x^4} - 6 \xi ^2 M\left(\int^1_0 \left(\frac{\partial u (x,t)}{\partial x} \right)^2 dx \right)\frac{\partial^2 u(x,t)}{\partial x^2} \\
    & = \frac{V^2}{(1-u(x,t))^2}
\end{split}   
\end{equation}
The four spatial boundary conditions for the fixed fixed beam are 
\begin{equation}
    u(0,t) = 0,~~ \frac{\partial u}{\partial x}(0,t) = 0,~~ u(1,t) = 0,~~ \frac{\partial u}{\partial x}(1,t) = 0
\end{equation}
\subsection{Determination of Dynamic pull-in parameters}
We express the normalized downward deflection of the microbeam, denoted as u(x, t), by approximating it using a known spatial function $\phi(x)$, with a time varying amplitude c(t). This mathematical representation can be expressed in a form that separates variables as follows
\begin{equation}
    u(x,t) = c(t)\phi(x)
\end{equation}
This is a single-mode approximation of the dynamic displacement of the beam. We use a static trial function, denoted by a subscript s, and a first mode shape of free vibration, denoted by subscript d as follows :
\begin{equation}
\begin{split}
 \phi_s(x) = (x^2-2x^3+x^4)
  ,~~~~
  \phi_d(x) = &cosh(\lambda x)-cos(\lambda x)\\- \frac{cos(\lambda)-cosh(\lambda)}{sin(\lambda)-sinh(\lambda)}&(sinh(\lambda x)-sin(\lambda x))
  \end{split}
\end{equation}
Value of $\lambda$ is taken as 4.73 for the fixed-fixed beam. We choose two different trial functions to check the dependence of solution on the trial functions. Our system is now a 1 DOF lumped parameter system. We can thus employ energy methods to find the unknown amplitude c(t), for pull-in parameter calculation. By previous studies, we know that at the dynamic pull-in point, there is no kinetic energy, and the supplied electrostatic energy equals the stored elastic energy in the beam. Appliyng this, we get the following energy equation:
\begin{equation}
    \begin{split}
         &\int _0 ^{\tilde c} \left[\int_0 ^1 c\left(\frac{d^2\phi}{dx^2}\right)^2 dx + [6\xi^2Mc^3]\left(\int_0 ^1 \left(\frac{d\phi}{dx}\right)^2\right)^2 \right] dc \\
    &= \int^{\tilde c} _0 \left[ (\tilde V)^2\int^1_0\frac{\phi dx}{(1-c\phi)^2}\right]dc
    \end{split}
\end{equation}
Here, $\tilde c$ is the maximum amplitude of the time varying amplitude c(t). We find it using the dynamic pull-in condition, which is
\begin{equation}
    \frac{d\tilde V}{d\tilde c} = 0
\end{equation}
We use the modulus ratio M = 1 for pull-in parameter calculation, for the study of narrow microbeams and for materials with low Poisson's ratio. We plot the variation in pull-in parameters with $\xi$ for an operating range of $0.4\leq \xi \leq 4$, for the two trial functions.
\begin{center}
\includegraphics[scale=0.6]{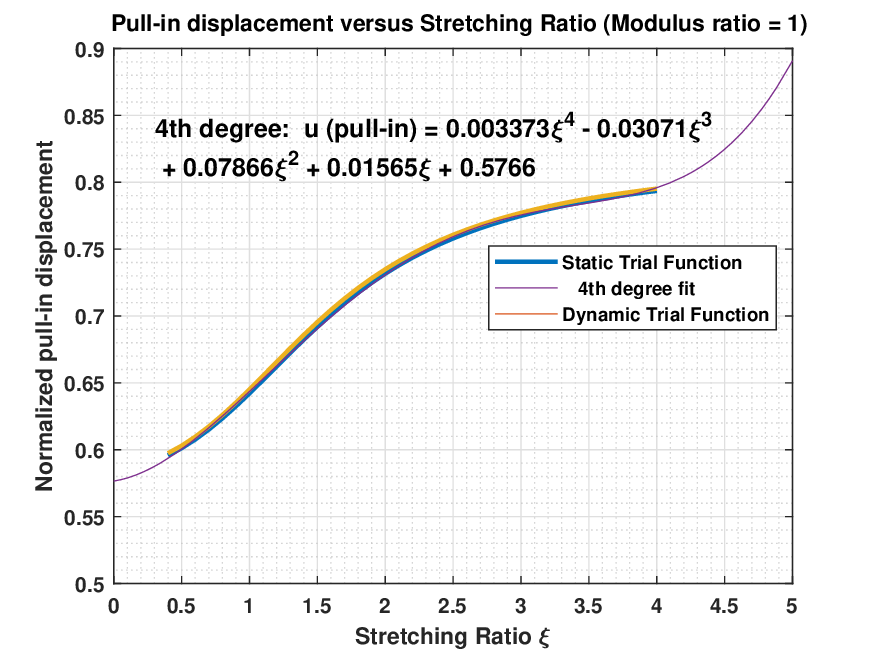}
\end{center}
\begin{center}
\includegraphics[scale=0.6]{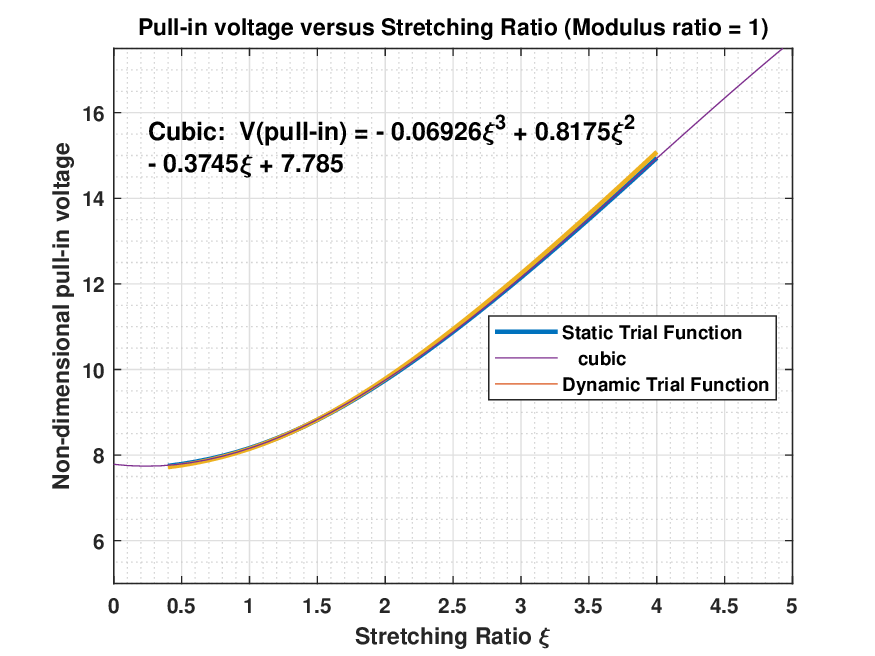}
\end{center}
The graphs for the two different functions are nearly identical, hence we use only the dynamic function for curve fitting and in the subsequent analysis. For both the pull-in parameters, there is a monotonic increase with $\xi$, suggesting that there is an increased stiffness of the beam with $\xi$, leading to larger voltages required for pull-in. This fact is emphasised by modelling the stretching effect as the stiffening parameter $\alpha$ in the parallel plates model, to increase the stiffness of the spring by adding a cubic term. The nature of graphs was also monotonic for that reduced order model. 
\subsection{Phase Plane Analysis}
We use the following energy equation to get velocity as a function of displacement to construct the phase-plane plots.
\begin{equation}
    \begin{split}
    &\left(\frac{1}{2} \int^1_0(\phi)^2dx \right)\left(\frac{dc_a}{dt}\right)^2 + \frac{c_a^2}{2}\int^1_0(\phi'')^2dx \\
    &+ c_a^4\left[\frac{3\xi^2M}{2}\right]\left(\int_0 ^1 \left(\frac{d\phi}{dx}\right)^2\right)^2 
    = \int^{c_a} _0 \left[ (V_a)^2\int^1_0\frac{\phi dx}{(1-c\phi)^2}\right]dc
    \end{split}
\end{equation}
        
In Eq.(19), $c_a$ is the corresponding maximum time amplitude for a given applied voltage $V_a$. The left side is the sum of the kinetic and stored elastic energies of the beam, and the right side is the supplied electrostatic energy. Thus, we obtain $dc_a/dt$ as a function of $c_a$ as given below:
\begin{equation}
    \begin{split}
    \left(\frac{dc_a}{dt}\right)^2 = &\left[\int^{c_a}_0 \left[(V_a)^2\int^1_0\frac{\phi \, dx}{(1-c\phi)^2}\right] \, dc - \frac{c_a^2}{2}\int^1_0 (\phi'')^2 \, dx \right. \\
    &\left. - c_a^4\left[\frac{3\xi^2M}{2}\int_0^1 \left(\frac{d\phi_d}{dx}\right)^2 \, dx \right] \left(\frac{1}{2}\int^1_0 \phi^2 \, dx \right)\right]
    \end{split}
\end{equation}

Eq.(20) is analogous to Eq.(8) obtained for the parallel plates model. Using this, we construct the phase plots for different $\xi$, as shown below, for a given applied voltage 10, similar to what was done for the plate model. We first find from the empirical estimates for pull-in voltage that 10 is a pull-in voltage for $\xi = 2.1 $.
\begin{center}
\includegraphics[scale=0.6]{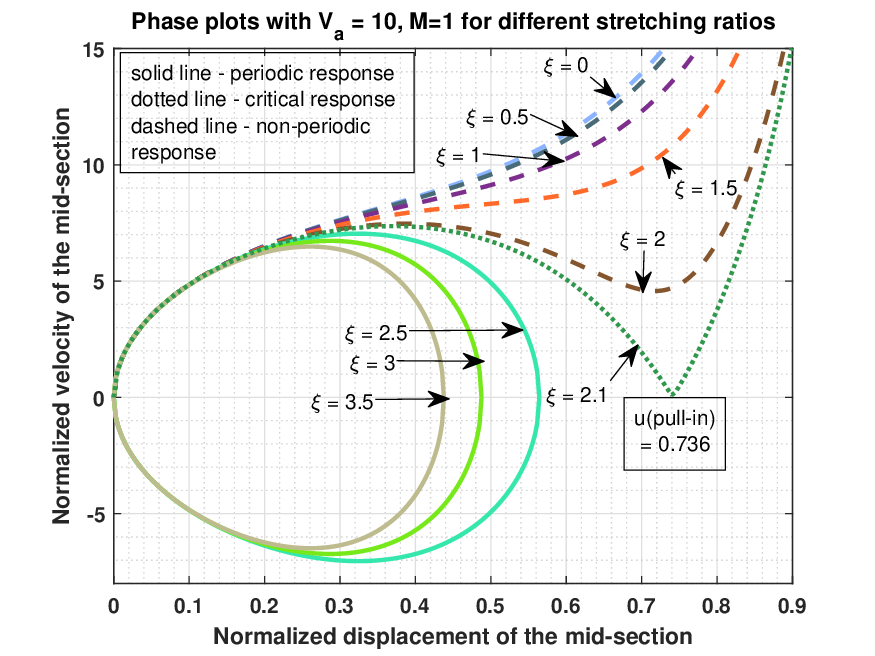}
\end{center}

The phase plot's behavior mirrors that observed in the parallel plates model. As the stretching effect intensifies with the parameter $\xi$, the response transitions from non-periodic to periodic. At the critical $\xi$ value corresponding to the pull-in voltage $V_a$, the phase plot briefly intersects the x-axis before rapidly diverging into an unbounded orbit. Notably, the periodic orbits, resembling ellipses, undergo compression or flattening at the extremity of displacement. This phenomenon is a consequence of the reduced maximum amplitude of the plates.

To elucidate this behavior, we resort to the analysis based on the parallel plate model. Two competing factors are at play: the electrostatic softening induced by the applied voltage and the beam's stiffness, which incorporates both a linear Young's modulus term and a cubic stretching effect term. As $\xi$ increases, the dominant stretching effect engenders a heightened effective stiffness, effectively constraining the beam to periodic orbits.

The subsequent section will delve into an in-depth exploration of the compression observed in the elliptical-like orbits. This will be accomplished through an analysis of time-measures specifically tailored for the fixed-fixed microbeam.
\subsection{Time Measures for the fixed-fixed microbeam}
We can use the same idea as for the reduced order model to calculate the time-estimates, by using the expression for $dc_a/dt$ obtained in Eq.(20). We find oscillation period for the periodic response and switching time for the non-periodic response as shown below:
\begin{equation}
    T_p = 2\int_0^{\bar c} \frac{dc_a}{(dc_a/dt)},~~
    T_{np} = \int_0^{1} \frac{dc_a}{(dc_a/dt)} 
\end{equation}
Here $\tilde c$ represents the maximum time amplitude of c(t) for the periodic response. We first plot the oscillation frequency variation with dimensionless applied voltage for a periodic response, for various values of $\xi$.
\begin{center}
\includegraphics[scale=0.6]{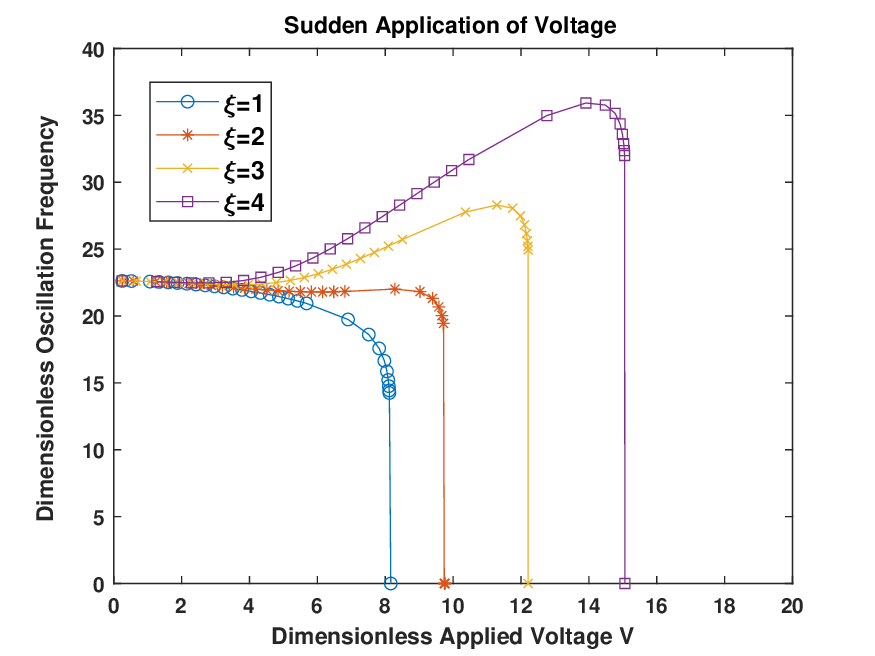}
\end{center}

The observed characteristics of the plots closely resemble those obtained for the parallel plates model. A sharp drop is observed near the pull-in voltage(indicated by a straight line to the pull-in voltage in the plots).In instances characterized by low values of $\xi$, the impact of stretching remains negligible, with the dominance of electrostatic softening effects. Consequently, an upsurge in the applied voltage leads to a discernible reduction in the oscillation frequency. This decline is attributed to the decreased effective stiffness, thereby prolonging the oscillation period and subsequently resulting in a lower frequency.\\

For intermediate values of $\xi$, approximately between 2 and 3, the frequency demonstrates relative stability across a wide spectrum of applied voltage. However, an observable increment is discernible as the applied voltage exceeds a certain threshold. This phenomenon can be rationalized by the association between higher voltages and amplified displacements, consequently leading to the prevailing influence of the cubic stretching term. %As a result, the overall rigidity of the beam experiences an augmentation, thereby accelerating the oscillatory behavior and subsequently elevating the frequency of oscillation. 
Consequently, this elucidates the notable rapidity of velocity changes proximate to the peak amplitude, as depicted in the preceding section's phase plots, which can be attributed to the corresponding surge in oscillation frequency.
%\section*{References}

% Please number citations consecutively within brackets \cite{b1}. The 
% sentence punctuation follows the bracket \cite{b2}. Refer simply to the reference 
% number, as in \cite{b3}---do not use ``Ref. \cite{b3}'' or ``reference \cite{b3}'' except at 
% the beginning of a sentence: ``Reference \cite{b3} was the first $\ldots$''

% Number footnotes separately in superscripts. Place the actual footnote at 
% the bottom of the column in which it was cited. Do not put footnotes in the 
% abstract or reference list. Use letters for table footnotes.

% Unless there are six authors or more give all authors' names; do not use 
% ``et al.''. Papers that have not been published, even if they have been 
% submitted for publication, should be cited as ``unpublished'' \cite{b4}. Papers 
% that have been accepted for publication should be cited as ``in press'' \cite{b5}. 
% Capitalize only the first word in a paper title, except for proper nouns and 
% element symbols.

% For papers published in translation journals, please give the English 
% citation first, followed by the original foreign-language citation \cite{b6}.

% \vspace{12pt}
% \color{red}
% IEEE conference templates contain guidance text for composing and formatting conference papers. Please ensure that all template text is removed from your conference paper prior to submission to the conference. Failure to remove the template text from your paper may result in your paper not being published.

\end{document}